\documentclass[12pt,a4paper]{article}
\usepackage{graphicx}
\usepackage{times}
\title{\bf  Spectroscopic Study of NGC 281 West} 
\author{Priya Hasan$^1$\thanks{e-mail:priya.hasan@gmail.com}
\vspace{1cm}\\
\normalsize $^1$ Maulana Azad National Urdu University, Hyderabad, India 500032}
\date{\mbox{}}
\begin{document}
\maketitle
\pagestyle{empty}
%
%
\def\bull{\vrule height .9ex width .8ex depth -.1ex}
\makeatletter
\def\ps@plain{\let\@mkboth\gobbletwo
\def\@oddhead{}\def\@oddfoot{\hfil\scriptsize\bull\quad
``First Belgo-Indian Network for Astronomy \& Astrophysics (BINA) workshop'', held in Nainital (India), 15-18 November 2016 \quad\bull}%
\def\@evenhead{}\let\@evenfoot\@oddfoot}
\makeatother
%
%
\def\beginrefer{\section*{References}%
\begin{quotation}\mbox{}\par}
\def\refer#1\par{{\setlength{\parindent}{-\leftmargin}\indent#1\par}}
\def\endrefer{\end{quotation}}
%
%

{\noindent\small{\bf Abstract:} 
NGC 281 is a complex region of star formation at 2.8 kpc. This complex is situated 300 pc above the Galactic plane, and appears to be part of a 270 pc diameter  ring of atomic and molecular clouds expanding at 22 km/s (Megeath et al. 2003).  It appears that two modes of triggered star formation are at work here: an initial supernova to trigger the ring complex and the initial O~stars and the subsequent triggering of  low mass star formation by photoevaporation driven molecular core compression. To get a complete census of the young stellar population, we use 
 Chandra ACIS 100 ksec coupled with data from 2MASS and Spitzer. The Master X-ray catalog  has  446 sources detected in different bandpasses.  We present the spatial distribution of Class~I, II and III sources to study the progress of star formation. We also determine the gas to dust ratio $N_H/A_K$ to be 1.93 $\pm$ 0.47 $\times 10^{22}$ cm$^{-2}$ mag$^{-1}$ for this region. 
{In this article, we present NGC 281 as a good target to study with the 3.6-m Devasthal Optical Telescope (DOT) in spectroscopy. With these spectra, we look for  evidence for the pre-main-sequence (PMS) nature of the objects, study the properties of the detected emission lines as a function of evolutionary class, and obtain spectral types for the observed young stellar objects (YSOs). The temperatures implied by the spectral types can be combined with luminosities determined from the near-infrared (NIR) photometry to construct Hertzsprung--Russell (HR) diagrams for the clusters. By comparing the positions of the YSOs in the HR diagrams with the PMS tracks, we can determine the ages of the embedded sources and study the relative ages of the YSOs with and without optically thick circumstellar disks.}
}
%
%

\section{Introduction}

Star formation can be initiated by a variety of processes: spontaneous gravitational instabilities in the stellar and gaseous medium, cloud collisions (especially in density-wave shocks) and triggered gravitational instabilities in compressed regions (like  spiral-arm dust lanes,  tidal arcs around interacting galaxies, shells and rings in galactic disks, and molecular clouds at the edges of HII regions) (Elmegreen 2011). 
{The spatial structure of young stars or protostars can lead to vital clues about the star formation sites and the motion of stars after they form}.

%

In the early stages of star formation, a protostar is shrouded in  circumstellar material and thus can only be observed as far-infrared (FIR) emission from the warm, dusty envelope (Class~0). This
envelope becomes less dense in Class~I objects. They often drive powerful outflows and create holes in their envelopes so radiation can escape. Eventually, the envelope disperses and
the star becomes visible as a classical T Tauri star (CTTS) or Class II source in infrared (IR) classification (Lada 1987, Lada \& Adams 1992). The disk still gives an IR excess over photospheric stellar emission. On the star itself, accretion shocks and coronal activity lead to X-ray emission (G\"unther et al. 2007; G\"unther 2012). Later, as the gas dissipates, the IR excess is not observed and the weak-line T Tauri stars (WTTS) cluster members cannot be distinguished from main sequence (MS) stars by IR observations. The young stars can be identified by their X-ray luminosity, which is far higher than for most older stars (Feigelson \& Montmerle 1999). Chandra X-ray data with its high resolution imaging is invaluable for a complete study of the young stellar population (Wolk et al. 2006, 2008). 
{Though a lot of studies have been made of NGC 281, to obtain the masses and ages of YSOs in NGC 281 West, spectroscopy of the target sources is necessary. Previous studies of NGC 281, do not include such studies of NGC 281 West. And hence here we present the science case of NGC 281 as a target for spectroscopic study.}

\section{NGC 281}
NGC 281 (the Pac Man Nebula) is a complex region of star formation. At a distance of about 2~kpc,  it lies a remarkable 300~pc above the galactic plane. It is generally divided into
2 subregions, east and west, based on the $^{12}$CO distribution  (Elmegreen \& Lada 1978). 

In the western part of NGC 281, the prominent HII region includes a young open cluster of stars  (IC 1590),  and a large lane of obscuring gas and dust. Within this lane are distinct CO and CS clumps. The open cluster of stars IC~1590 visible around the center is about 3.5$\times 10^6$ years old (Guetter \& Turner 1997).  At the heart  of IC~1590 is the trapezium-like HD 5005 containing an O6.5,  an O8, and an O9 star. Guetter \& Turner (1997) identified 61 cluster members down to early K types.  They derived a radius of 4$^\prime$ and an initial mass function with $\alpha \sim 2 $. This is somewhat flatter than other clusters and may indicate a missing population of stars. Within the dust lane lies a large molecular cloud about 15 pc in diameter with a  M$_{VIR}\sim 3.1\times 10^4 M_{\odot}$.  The cloud seems to be under active
stress from the HII region.  The peak of the molecular emission occurs at the cloud edge close to a water maser and an IRAS point source.  The molecular emission terminates sharply along the edge of the dust lane.  Megeath \& Wilson (1997) mapped the interface and found 2 condensations of C$^{34}$S (3-2) arising from highly compressed gas suggestive of the recent passage of a shock. 

To the east, along the ionization front, is a smaller, less stressed, embedded cluster.
The brightest source in this region is coincident with a molecular outflow. This is part of the large molecular cloud  about 18 pc in diameter with a M$_{VIR}\sim 1.5\times 10^4M_{\odot}$.
Henning et al. (1994) detected  a small, highly reddened cluster toward the northern
end. 
 
 Megeath et al. (2003) found that NGC 281 is the diamond atop a 270 pc diameter ring of atomic and molecular clouds.  They suggest that two modes of triggered star formation are at work.  An initial supernova to trigger the ring complex and the initial O stars and the subsequent triggering of  low mass star formation by photoevaporation driven molecular core compression.

Sharma et al. (2012) presented a multiwavelength study of the NGC~281 complex. They used deep wide-field optical UBVIc  photometry, slitless spectroscopy and  the IRAC  and Chandra data.  They suggest triggered star formation at the periphery of the cluster region based on the radial distribution of the young stellar objects, their ages, $\Delta$ (H- K) NIR-excess, and the fraction of classical T Tauri stars. 

\section{Observations  \& Reduction}
\begin{figure} 
\centering
\resizebox{0.45\textwidth}{!}{\includegraphics{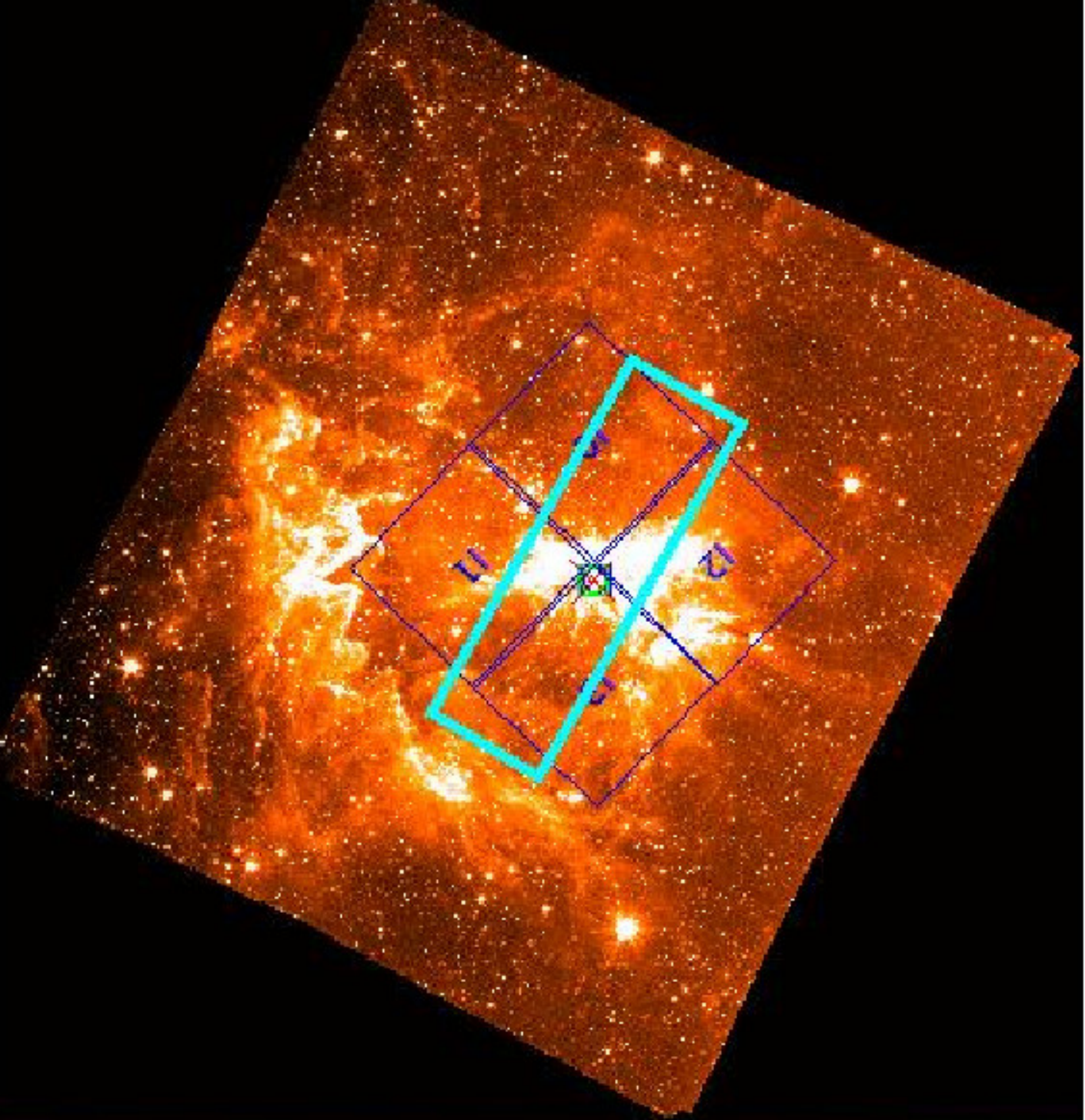}}
\caption{The (3.6 $\mu$m) Spitzer image of NGC 281 from the Warm Spitzer image. The large blue square is the Chandra ACIS field and the cyan box is the Spitzer cyro data.}
\label{ngc281f}
\end{figure}

To visualize the data used in this paper, Fig. \ref{ngc281f} shows the IRAC (3.6 $\mu$m)  image of NGC 281 from the Warm Spitzer data. The large blue square is the Chandra ACIS field and the smaller cyan box is the Spitzer cyro data.

\subsection{X-ray}
NGC 281 was observed with Chandra in 2005 for almost 100 ks with the ACIS  in  three sets of observations and downloaded from the Chandra Data Archive.\footnote{See  http://cxc.harvard.edu/cda/} Table \ref{xobs} lists the observations, their durations and details. 

\begin{figure} 
\centering
\resizebox{0.6\textwidth}{!}{\includegraphics{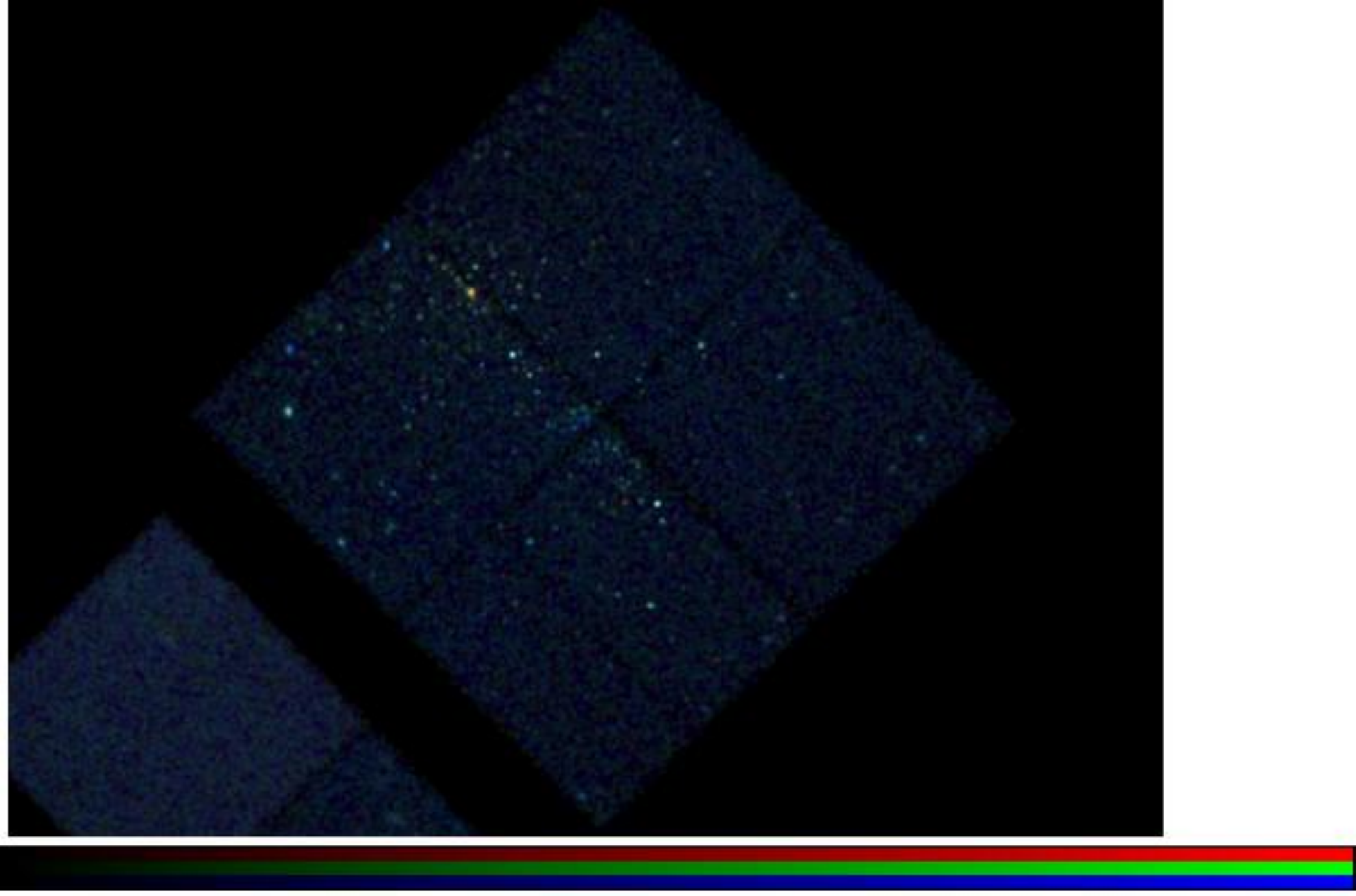}}
\caption{NGC 281: RGB color-coded image in X-rays.}
\label{xray}
\end{figure}

\begin{table}
\caption{Chandra observations of NGC 281} 
\label{xobs}
\small
\begin{center} 
\begin{tabular}{| l | l| l| l |}
\hline

Obsid/PID  &  Start Time      &    Duration  &      Roll Angle\\
Chandra    &                  &         (sec)  &      (degrees   \\
           &                  &              &    W. of North) \\ \hline
7206       &  2005-11-08T13:42:50 & 23,200   &        134.5   \\
5424       &  2005-11-10T18:20:31 & 62,700   &        134.5  \\
7205       &  2005-11-12T22:41:58 &  13,200  &        134.5   \\
\hline
\end{tabular}
\end{center} 
\end{table}

The merged dataset was  used to run WavDetect (Freeman et al. 2001) to detect point sources in the soft, medium, hard and broad bands using a threshold value of 10$^{-5}$.  Fig \ref{xray} shows an X-ray RGB image obtained by using soft, medium and hard X-ray sources. We generated a  master catalog of X-ray sources matched at 1" with atleast one detection in any band with  source significance (given by WavDetect)$>$ 2.75. The master X-ray catalog now has  446 sources: 145 soft, 213 medium, 154 hard, 303 broad. Of these, 26 are far  off-axis and hence will be ignored. So the final catalog now has 420 sources.

\subsection{Near Infrared Data}
\begin{figure} 
\centering
\resizebox{0.55\textwidth}{!}{\includegraphics{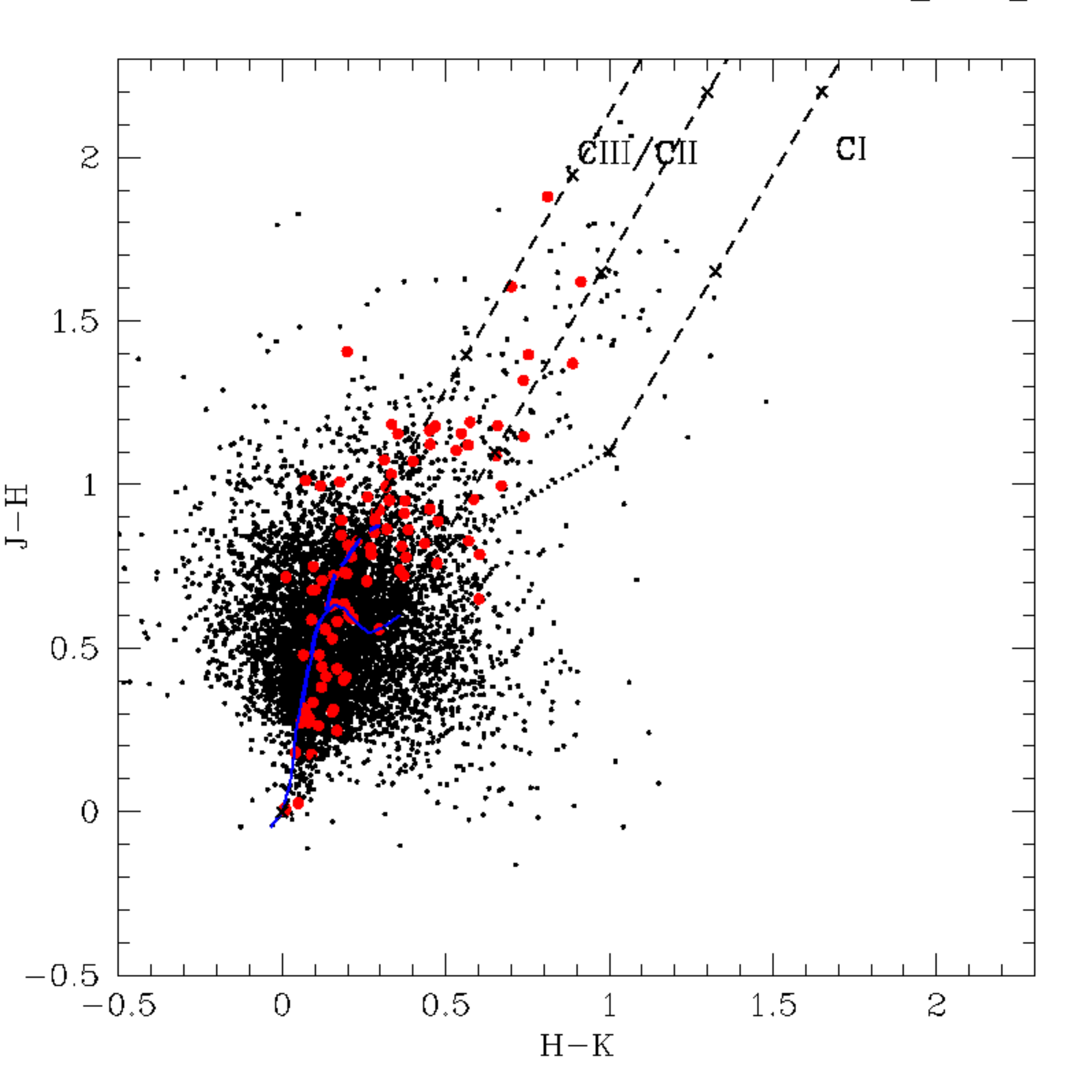}}
\caption{Color-Color diagram for stars with $\sigma < 0.1$ in  NGC 281, the X-ray matched stars are shown in red.}
\label{jhk}
\end{figure}

In the NIR we use J, H, and Ks photometry obtained from the 2MASS  point source catalog (Skrutskie et al. 2006). 3608 stars have data with errors in $JHK$ data  $<$ 0.1. Of these only 100  matched  with our X-ray catalog and are marked in  Fig. \ref{jhk}. 

\begin{table}
\caption{Spitzer Cyro data of NGC 281 West}
\label{sobs}
\small
\begin{center} 
\begin{tabular}{| l |l| l | l |}
\hline
Channel &  Central wavelength&   No of stars   &       No of stars with sigma$<$0.2 \\
        &    &   with data   &                            \\ 
\hline
Ch1   & 3.6 $\mu$m &     1571(63\%)  & 1298(52\%)  \\
Ch2   & 4.5 $\mu$m &        1557(63\%)  & 1277(52\%)  \\
Ch3   & 5.8 $\mu$m &         389(16\%)  & 265(11\%)  \\
Ch4   & 8 $\mu$m   &     196(8\%)    & 111(4\%)  \\
Ch24  & 24 $\mu$m  &       31(1\%)    & 26(1\%) \\

\hline
\end{tabular}
\end{center}
\end{table}

\subsection{Spitzer}       

{Data were obtained by the Spitzer Space Telescope with the Infrared Array Camera (IRAC; Fazio et al. 2004) and by the Multiband Imaging Photometer for Spitzer (MIPS; Rieke et al. 2004) in separate programs. The 24 $\mu$m image was originally taken as part of Spitzer program PID 20726 (J. Hester PI) following up on HST observations of HII Regions. The IRAC data were taken during the warm Spitzer period and hence only include Ch1 = 3.6 $\mu$m, Ch2 = 4.5 $\mu$m (PID 60078).}

\subsubsection{IRAC Cryo}
The Spitzer Cyro data has data at  Ch1 = 3.6 $\mu$m, Ch2 = 4.5 $\mu$m, Ch3 = 5.8 $\mu$m and Ch4 = 8 $\mu$m  with 2475 sources. Details are in Table \ref{sobs}.

\begin{figure}
\begin{minipage}{8cm}
\centering
\includegraphics[width=8.5cm]{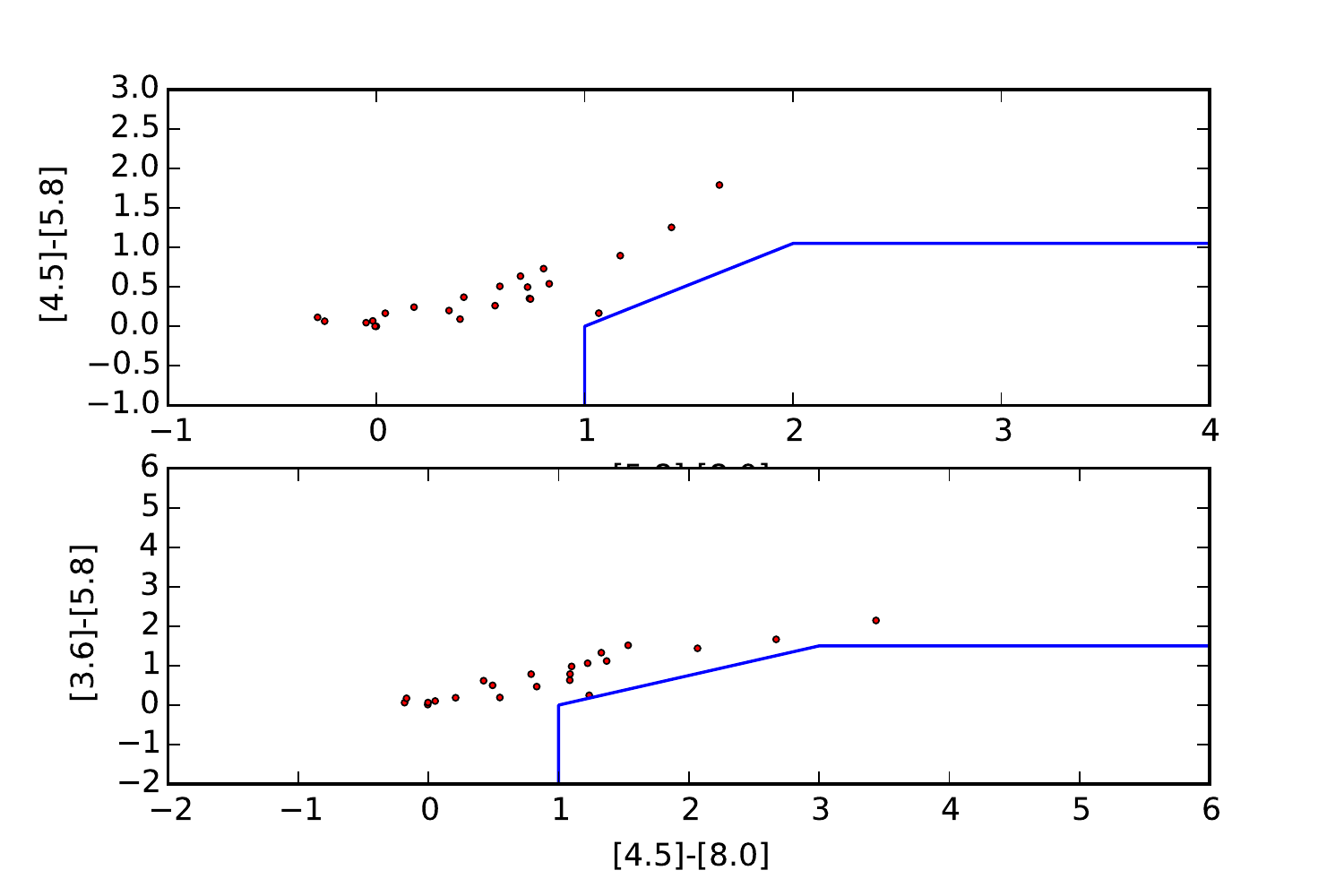}
\caption{Isolation of unresolved star forming galaxies using color-color plots using the criteria described by Gutermuth et al. (2009).\label{g13}} 
\end{minipage}
\hfill
\begin{minipage}{8cm} 
\centering
\includegraphics[width=8.5cm]{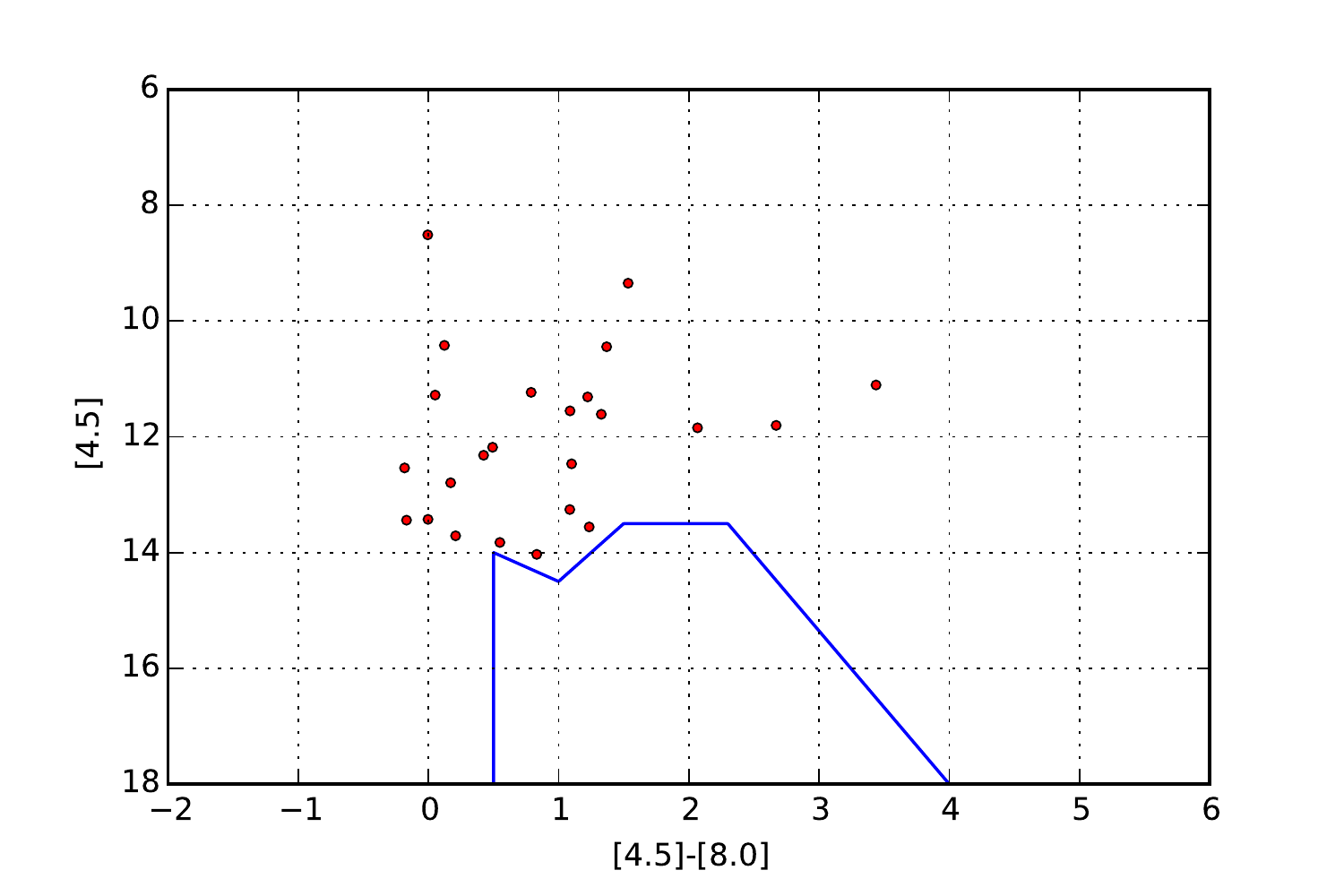}
\caption{Isolation of broad AGNs using the [4.5] versus [4.5]-[8.0] color-magnitude diagram  and the criteria described by Gutermuth et al. (2009).\label{g14}}
\end{minipage}
\end{figure}

\begin{figure}
\begin{minipage}{8cm}
\centering
\includegraphics[width=8.5cm]{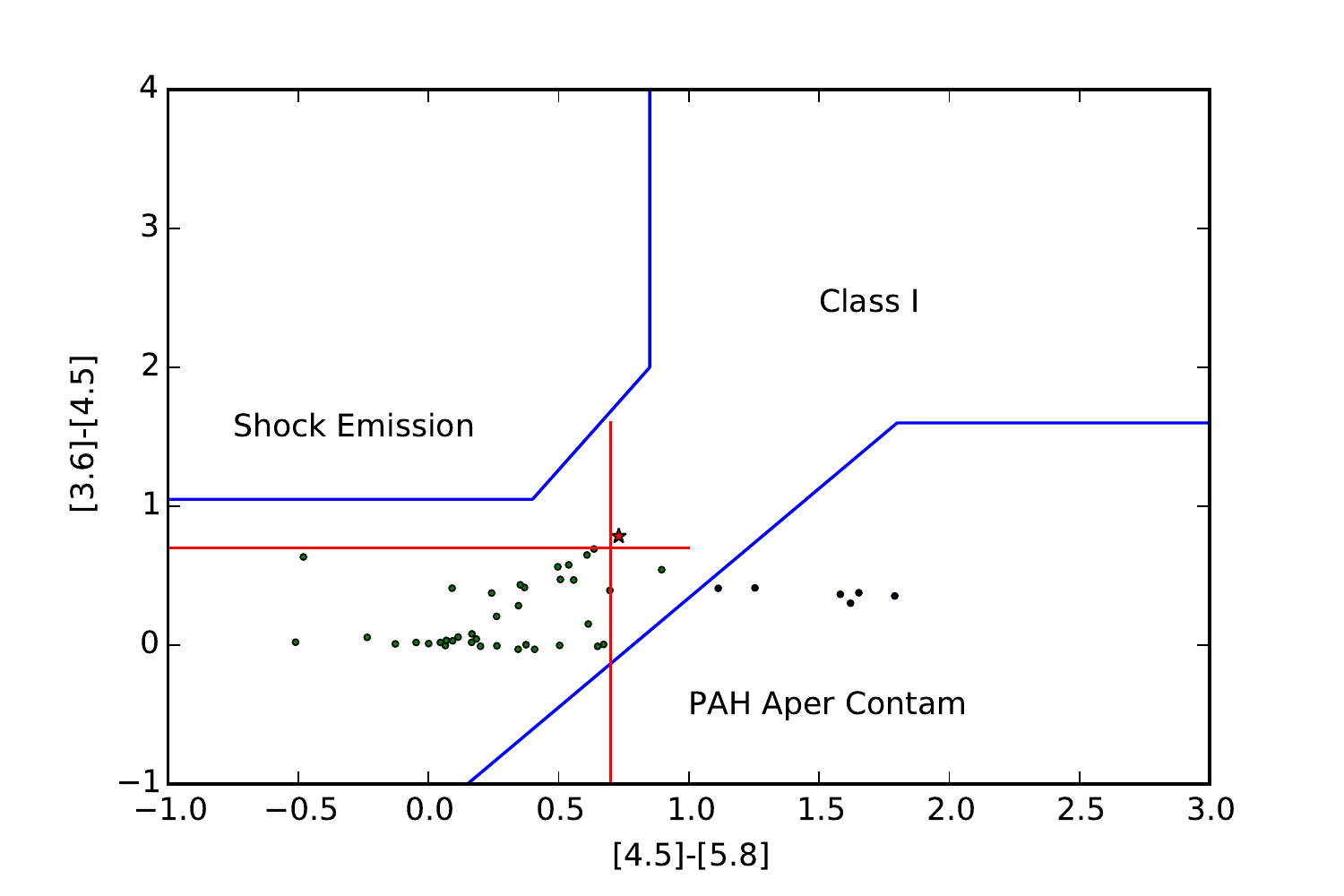}
\caption{The [3.5]-[4.5] versus [4.5]-[5.8] color-color diagram  for the isolation of unresolved shock emission sources and PAH aper contamination using the criteria described by Gutermuth et al. (2009). One Class I source (red) is identified (red and blue line). \label{g15}}
\end{minipage}
\hfill
\begin{minipage}{8cm}
\centering
\includegraphics[width=8.5cm]{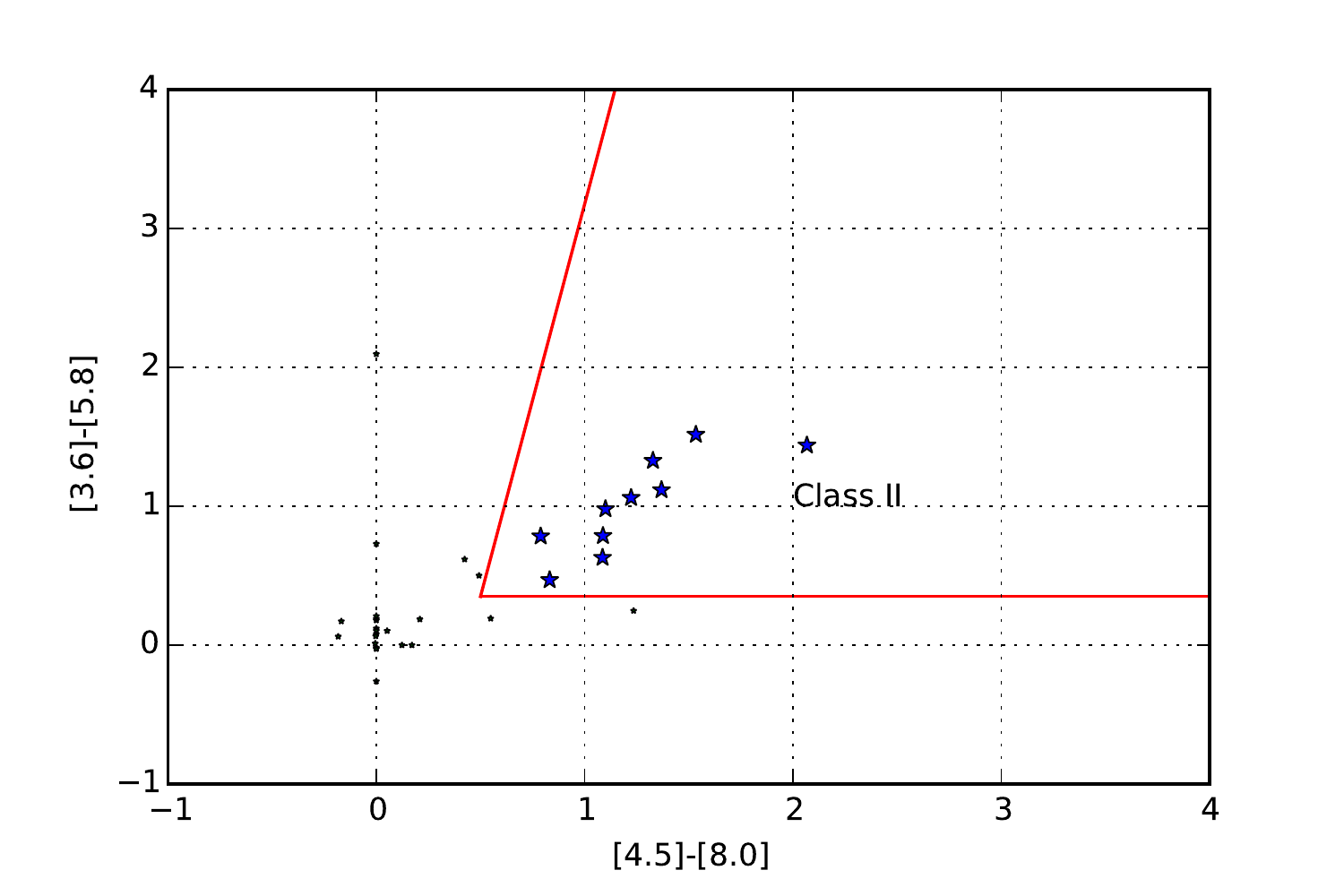}
\caption{Color-color diagram for the isolation of Class II (blue) sources  using the criteria described by Gutermuth et al. (2009) (red line). \label{g16}}
\end{minipage} 
\end{figure}

{To identify possible YSOs, we used three methods: selection of stars with IR
excesses on IR color-color diagrams,  identification of X-ray sources with IR detections and finally a search for extremely red mid-infrared (MIR) sources among the detections using 24 $\mu$m MIPS data. We did the YSO classification based on Guthermuth et al. (2009) 4-band IRAC data   for the decontamination and classification of Class I and Class II sources (see Figs \ref{g13} to \ref{g16}).  Fig. \ref{g13} was for the isolation of unresolved star forming galaxies. Fig. \ref{g14} was for the isolation of broad active galactic nuclei (AGN) using color-magnitude diagram. Figure \ref{g15} was for the isolation of unresolved shock emission sources and polycylic aromatic hydrocarbons (PAH)  contamination and shows the  Class I  sources (red). Fig. \ref{g16} shows  the identification of  Class II (blue) sources.} 

\subsubsection{Results of Warm mission data}

\begin{figure} 
\centering
\resizebox{0.6\textwidth}{!}{\includegraphics{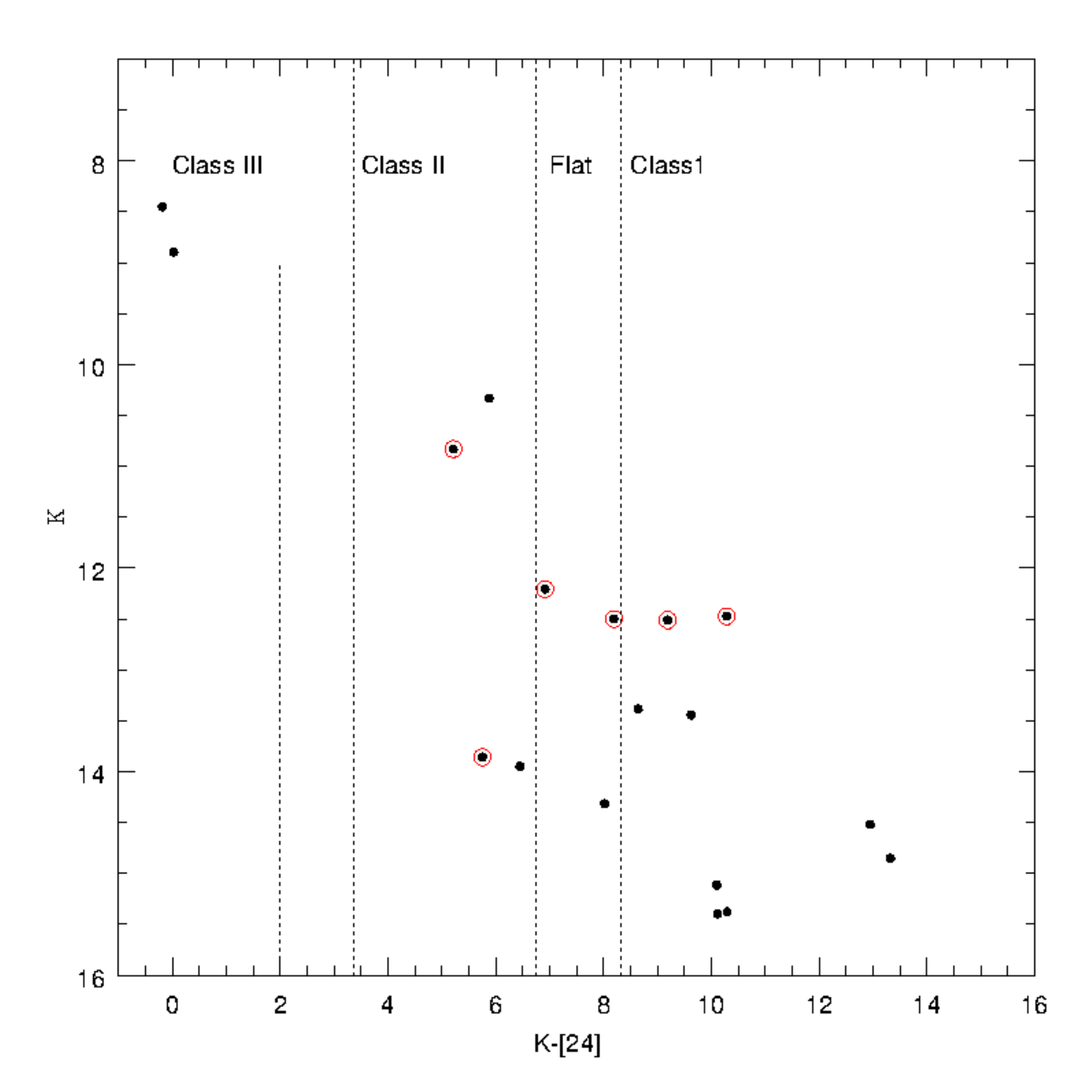}}
\caption{Classification based on $K$ versus $K-[24]$ color-magnitude diagram (Rebull et al. 2007).  The black dots show 18 IR sources with $K-[24]$ data, classified as Class I (9), Flat (3), Class II (4) and  Class III (2). The 6 objects that are also X-ray sources are shown in red circles and include Class I(2), Flat Spectrum (2) and Class II (2) sources,  but no Class III sources.}
\label{r1}
\end{figure}

Spitzer Warm mission data (references described earlier) is available for a larger field at 3.6 and 4.5 $\mu$m.  It has 30001 stars and 164 matches with our X-ray catalog. For the NGC 281 West area, we classified stars into Class I, Flat, Class II, and Class III objects, using the $K$ versus $K-[24]$ color-magnitude diagram , as in Rebull et al. (2007) (Fig. \ref{r1}). Objects with  $K-[24]$ above 8.31 are Class I objects, between 6.75 and 8.31 are flat-spectrum objects,  between 3.37 and 6.75 are Class II objects, and below 3.37 are Class III.  We have 18 IR sources with $K-[24]$ data, and thus classified Class I (9), Flat (3), Class II (4)and Class III (2) sources, shown in black. The 6 objects that are also X-ray sources are shown in red circles and include Class I(2), Flat Spectrum (2) and Class II (2) sources,  but no Class III sources.(Fig. \ref{r1}).

\subsection{X-ray spectral analysis} 
{The analysis of the X-ray spectrum of each source was performed to determine the bulk temperature of the corona ($kT$) and the intervening column of hydrogen ($N_H$). For each source with over 30 counts, the source and background pulse height distributions in the total band (0.3-8.0~keV) were constructed}. We used the CXC SHERPA program to generate spectral fits and to solve simultaneously for $N_H$ and $kT$.

The absorption of X-rays takes place due to metal atoms and ions along the line of sight and gets measured by $N_H$. The dust causes extinction of the optical and IR light  and is characterised by $A_V$ and $A_K$, respectively. We  obtained the gas to dust ratio along the line of sight as $N_H/A_K$ to be 1.93 $\pm$ 0.47 $\times 10^{22}$ cm$^{-2}$ mag$^{-1}$, in agreement with the the standard interstellar medium  gas to dust ratio of $N_H = 1.6 \times 10^{22} A_K$ (Vuong et al. 2003).

\section{Discussion}
\begin{figure} 
\centering
\resizebox{0.7\textwidth}{!}{\includegraphics{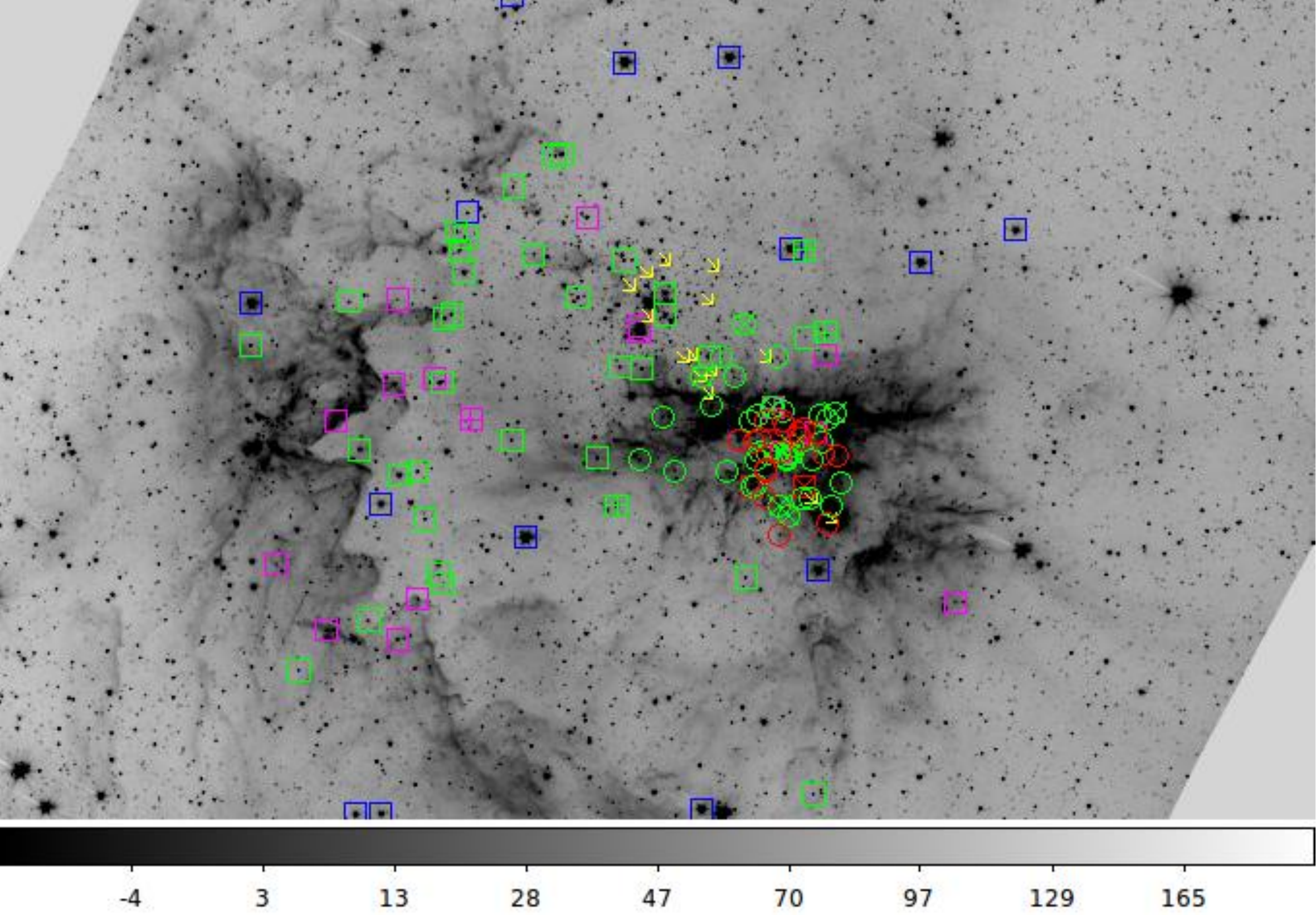}}
\caption{Distribution of sources in NGC 281. The Class I , II and III and Flat spectrum sources are shown in red, green, blue and magenta. The ones with X-ray detections are shown with crosses. Circles are used for stars classified with the Guthermuth et al. (2009) method while those classified with the Rebull et al. (2007) classification method are given with squares.}
\label{dis}
\end{figure}

%
We found 420 X-ray sources  with a source  significance above 2.75. There are 196 matches of X-ray and warm Spitzer  data sources and 104 stars matched with NGC 281 West cyro data. We defined these to be YSO/PMS candidates and classified them using the Guthermuth et al. (2009) and Rebull et al. (2007) classification scheme. The distribution of these sources are given in Fig. \ref{dis}. We clearly see that the Class I sources are more concentrated and the Class II and Class III sources are more dispersed. 

For sources with more than 30 counts, we did X-ray spectroscopy to find $N_H$ and $kT$. We also found the $A_K$ value and thus obtained the gas to dust ratio $N_H/A_K$ to be 1.93 $\pm$ 0.47 $\times 10^{22}$ cm$^{-2}$ mag$^{-1}$, in agreement with the standard interstellar ratio. 

These are preliminary results and a complete paper will be published soon.

\section{Conclusion}

 NGC 281 would be an ideal case study for the 3.6-m Devasthal Optical Telescope (DOT) using spectroscopy. 
 {We propose to take spectra of  YSOs identified in previous Spitzer and Chandra observations.  With these spectra we can search for  evidence for the PMS nature of the objects and study the properties of the detected emission lines as a function of evolutionary class, and obtain spectral types for the observed YSOs. The temperatures obtained by the spectral types can be combined with luminosities determined from the NIR photometry to construct HR diagrams for the cluster.  The HR diagrams can be used to determine the ages of the embedded sources and study the relative ages of the YSOs with and without optically thick circumstellar disks. We can also examine the spatial distribution and extinction of the YSOs as a function of their isochronal ages (Winston et al. 2009). 
The spectral class can be used in combination with NIR photometry and
PMS models to determine ages and masses for the YSOs. These measured ages and masses can be used to determine the initial mass functions of embedded clusters, study the star formation history of the clusters, and examine the evolution of circumstellar disks as a function of age.}

\section{Acknowledgements}
The author would like to thank Scott Wolk under whose guidance this work was done and Hans Moritz G\"unther and Elaine Winston whose help has been very useful. The author would also like to thank IAU-OAD and especially Katrien Kolenberg, who funded her visit to the Harvard-Smithsonian Center for Astrophysics (CfA), USA where this work was carried out.

The scientific results reported in this article are based  on  data obtained from the Chandra Data Archive. This research has made use of software provided by the Chandra X-ray Center (CXC) in the application packages CIAO, ChIPS, and Sherpa. This work is partly based on observations made with the Spitzer Space Telescope,  operated by the Jet Propulsion Laboratory, California Institute of Technology under a contract with NASA. This publication makes use of data products from the Two Micron All Sky Survey, which is a joint project of the University of Massachusetts and the Infrared Processing and Analysis Center/California Institute of Technology, funded by the National Aeronautics and Space Administration and the National Science Foundation.

\footnotesize
\beginrefer

\refer  Elmegreen B. G. 2011, EAS Publications Series, 51, 45

\refer  Elmegreen B.G, Lada C.J. 1978, ApJ, 219, 467

\refer  Fazio G. G., Hora J. L., Allen L. E. et al. 2004, ApJS, 154, 10

\refer  Feigelson E. D., Montmerle T. 1999, ARAA, 37, 363

\refer  Freeman P.E., Doe S., Siemiginowska A. 2001, SPIE Proceedings, 4477, 76

\refer  G\"unther H. M., Schmitt J. H. M. M., Robrade J., Liefke C. 2007, A\& A, 466, 1111
	
\refer  G\"unther H. M., Wolk S. J., Spitzbart B. et al. 2012, AJ, 144,  4

\refer  Guetter H.H., Turner D. G. 1996, AJ, 113, 2116

\refer  Gutermuth R. A., Megeath S. T., Myers P. C. et al. 2009, ApJS, 184, 18

\refer  Henning T., Martin K., Reimann H.-G. et al. 1994, A\& A, 288, 282

\refer  Lada C. J. 1987, in IAU Symposium,  Vol. 115, \textit{Star Forming Regions}, Peimbert M., Jugaku J., eds., 1

\refer  Lada C. J., Adams F. C. 1992, ApJ, 393, 278 

\refer  Megeath S. T., Biller B., Dame T. M. et al. 2003, RMxAC, 15, 151

\refer  Megeath S. T., Wilson T.L. 1997, AJ, 114, 3

\refer  Rebull L. M., Stapelfeldt K. R., Evans II N. J. et al. 2007, ApJS, 171, 447

\refer  Rieke G. H., Young E. T., Engelbracht C. W. et al. 2004, ApJS, 154, 25

\refer  Sharma S., Pandey A.K., Pandey J. C. et al. 2007, MNRAS, 380, 1141

\refer  Skrutskie M.F., Cutri R. M., Stiening R. et al. 2006, AJ, 131, 1163

\refer  Vuong M. H., Montmerle T., Grosso N. et al. 2003, A\&A, 408, 581-599

\refer  Winston E., Megeath S. T., Wolk S. J. et al. 2009, AJ, 137, 6, 4777

\refer  Wolk S.J., Spitzbart B.D., Bourke T.L. et al. 2006, AJ, 132, 1100

\refer  Wolk S.J., Spitzbart B.D., Bourke T.L. et al. 2008, AJ, 135, 693

\endrefer
    
\end{document}